\documentclass[11pt]{article}

\usepackage{times}

\usepackage{amssymb}

\usepackage{epsfig}

\renewcommand*{\Pr}{\mathop{\mathrm{Pr}}}

\def \qedbox{\hfill\vbox{\hrule\hbox{\vrule height1.3ex\hskip0.8ex\vrule}\hrule}}

  \newcommand{\goesto}{\rightarrow}
  \newtheorem{theorem}{Theorem}
 \newtheorem{remark}{Remark}
\newtheorem{observation}{Observation}
  \newtheorem{definition}{Definition}
  \newtheorem{proposition}{Proposition}
\newtheorem{corollary}{Corollary}

\def\AND{\wedge}
\def\OR{\vee}

\def\goesto{\rightarrow}

\newtheorem{lemma}{Lemma}[section]

\newtheorem{example}{Example}

\newtheorem{claim}{Claim}

\def\qed{\hfill$\Box$\newline\vspace{5mm}}

\begin{document}

\pagestyle{empty}

\title{Coarse and Sharp Thresholds of Boolean Constraint Satisfaction Problems\footnote{This work has been supported by the Department of Energy under contract W-705-ENG-36}}
\author{Gabriel Istrate\thanks{e-mail: istrate@lanl.gov,
        CCS-5, Basic and Applied Simulation Science,
        Los Alamos National Laboratory,
        Mail Stop M 997, Los Alamos, NM 87545, U.S.A.}}
\date{}

\maketitle
\thispagestyle{empty}

\section*{Abstract}

We study threshold properties of random constraint satisfaction problems under 
a probabilistic model due to Molloy \cite{molloy-stoc2002}. We give a sufficient condition for the existence of a sharp threshold that leads (for boolean constraints) to a necessary and sufficient for the existence of a {\em sharp threshold} in the case where constraint templates are applied with equal probability,
solving thus an open problem from \cite{creignou:daude:sat2002}.

\section{Introduction}

Classifying threshold properties of random constraint satisfaction
problems is a problem that has been intensely studied in recent
literature. The well-known Friedgut-Bourgain result
\cite{friedgut:k:sat} showed that 3-satisfiability has a sharp
threshold, via a more general result on threshold properties of
monotonic sets.

For random satisfiability problems (i.e. constraint satisfaction
problems with a boolean domain) we have obtained
\cite{istrate:ccc00} a classification of thresholds of such
properties, under a random model that allows constraints of
different lengths. The results from \cite{istrate:ccc00}
intuitively show that satisfiability problems with a coarse
threshold qualitatively ``behave like random Horn
satisfiability'', a problem with a known coarse threshold
\cite{istrate:horn}. A drawback of these results is that the
classification is not ``structural'' (that is defined in terms of
properties of the formula hypergraph), and thus does not offer
suggestions for a suitable generalization to non-boolean domains.

Recently Molloy \cite{molloy-stoc2002} has investigated a model of
random constraint satisfaction that allows constraints of the same
arity  and unequal probabilities for the application of the given
constraint templates (hence is only partly comparable with the
results in \cite{istrate:ccc00}). He offers a nice structural
condition that is necessary and sufficient for the existence of a
weaker form of threshold property (he calls  a {\em transition}).

While Molloy's model certainly has several remarkable features
(such as the location of the transition in the region of formulas
whose ratio between clauses and variables is  constant), it was
observed in \cite{molloy-stoc2002} that the necessary and
sufficient condition for the existence of a transition is {\em
not} sufficient for the existence of a {\em sharp} threshold. The
counterexample involves binary constraints on $\{0,1,2,3\}$ and a
nonuniform probability distribution on the set of such templates.
For the case of a uniform probability distribution he gave
(Theorem 6) a sufficient condition for the existence of a sharp
threshold.

In this note we strengthen this latter result: We give a more general sufficient condition for the existence of a sharp thresholds, that allows
to obtain a precise classification of sharp/coarse thresholds for
random {\em satisfiability problems}
(problems with a binary domain), confirming thus a conjecture due to Creignou and Daud\'{e}. The key to these results is a ``monotonicity'' property of the
Friedgut-Bourgain method for proving the existence of a sharp
threshold. We have first observed this property in
\cite{istrate:ccc00}, and it is, we believe, of independent
interest.

\section{Preliminaries}

Throughout the paper we will assume familiarity with the general
concepts of phase transitions in combinatorial problems (see e.g.
\cite{martin:monasson:zecchina}), random structures
\cite{bol:b:random-graphs}. Some papers whose concepts and methods
we use in detail (and we assume greater familiarity with) include
\cite{friedgut:k:sat,achlioptas:friedgut:kcol}.

Consider a monotonically increasing problem $A=(A_{n})$. The two models used in the theory of random structures are:
\begin{itemize}
\item The {\em constant probability model} $\Gamma(n,p)$. A random sample from this model is obtained by independently
setting to 1 with probability $p$ each bit of the random string.
\item The {\em counting model} $\Gamma(n,M)$. A random sample from this model is obtained by setting to 1 $M$ different bits chosen uniformly at random from the $n$ bits of the random sample.
\end{itemize}

As
usual, for $\epsilon >0$ let $p_{\epsilon}= p_{\epsilon}(n)$
define the canonical probability such that $\Pr_{x \in
\Gamma(n,p_{\epsilon}(n))}[x \in A]= \epsilon$. The probability that a random sample $x$ satisfies property $A$
(i.e. $x\in A$) is a monotonically increasing function of $p$.
{\em Sharp thresholds} are those for which this function has a
``sudden jump'' from value 0 to 1:

\begin{definition} \label{sharp}
Problem $A$ has a {\em sharp threshold} iff for every $0<\epsilon
< 1/2$, we have $\lim_{n\goesto \infty} \frac{p_{1-\epsilon}(n)-
p_{\epsilon}(n)}{p_{1/2}(n)} = 0$. $A$ has {\em a coarse
threshold} if for some $\epsilon > 0$ it holds that
$\underline{\lim}_{n\goesto \infty} \frac{p_{1-\epsilon}(n)-
p_{\epsilon}(n)}{p_{1/2}(n)} > 0$.
\end{definition}

For satisfiability problems (whose complements are monotonically
increasing) the constant probability model amounts to adding every
constraint (among those allowed by the syntactic specification of
the model) to the random formula independently with probability
$p$. Related definitions can be given for the other two models for
generating random structures, the {\em counting model} and {\em
the multiset model} \cite{bol:b:random-graphs}. Under reasonable
conditions \cite{bol:b:random-graphs} these models are equivalent,
and we will liberally switch between them. In particular, for
satisfiability problem $A$, and an instance $\Phi$ of $A$,
$c_{A}(\Phi)$ will denote its {\em constraint density}, the ratio
between the number of clauses and the number of variables of
$\Phi$. To specify the random model in this latter cases we have
to specify the constraint density as a function of $n$, the number
of variables. We will use  $c_{A}$ to denote the value of the
constraint density $c_{A}(\Phi)$ (in the counting/multiset models)
corresponding to taking $p=p_{1/2}$ in the constant probability
model. $c_{A}$ is a function on $n$ that is believed to tend to a
constant limit as $n\goesto \infty$. However, Friedgut's proof
\cite{friedgut:k:sat} of a sharp threshold in $k$-SAT (and our
results) leave this issue open.

\begin{definition}\label{model} Let ${\cal D} = \{0,1,\ldots, t-1\}$, $t\geq 2$ be a fixed set.
Consider the set of all $2^{t^{k}}-1$ potential nonempty binary
constraints on $k$ variables $X_{1}, \ldots, X_{k}$. We fix a
set of constraints ${\cal C}$ and define the random model $CSP({\cal C})$.
A random formula from $CSP_{n,p}({\cal C})$ is specified by the
following procedure:
\begin{itemize}
\item $n$ is the number of variables.
\item for each $k$-tuple of {\em ordered} distinct variables $(x_{1}, \ldots, x_{k})$ and
each $C\in {\cal C}$ add constraint $C(x_{1}, \ldots, x_{k})$ independently with probability $p$.
\end{itemize}

When the number of variables $n$ is known (or unimportant) we will
drop it from our notation, and write $CSP_{p}({\cal C})$ instead.

\end{definition}

\begin{remark}
The model in Definition~\ref{model} differs from the model in \cite{molloy-stoc2002} in two respects:
\begin{itemize}
\item It is a ``constant probability''-type model (the one in
\cite{molloy-stoc2002} is a ``counting''-type model). \item More
importantly, all templates $C\in {\cal C}$ are applied with the
same probability (in \cite{molloy-stoc2002} a probability ${\cal
P}$ on the set of all templates is considered, and templates are
instantiated according to this distribution).
\end{itemize}
However, as discussed in \cite{molloy-stoc2002} (Remark 2) one can map Molloy's model onto a modified version of the
constant probability model. In particular Definition~\ref{model} corresponds to this mapping when ${\cal P}$ is the uniform distribution,
which is why we chose this definition as the starting point.
\end{remark}

\begin{definition}
If $\Phi$ is a constraint satisfaction problem, the {\em formula hypergraph}
of $\Phi$ is the hypergraph $H$ that has
\begin{itemize}
\item the set of variables that appear in $\Phi$ as vertices.
\item the sets of variables that appear together in a constraint of $\Phi$ as edges.
\end{itemize}

$H$ is {\em tree-like} if it is a connected acyclic hypergraph, and is {\em
unicyclic} if there exists an edge $e\in H$ such that $H\setminus e$ is tree-like.

\end{definition}

\section{Coarse and sharp thresholds of random generalized constraint satisfaction problems}

In this section we study the sharpness of the threshold for random
generalized constraint satisfaction problem defined by Molloy
\cite{molloy-stoc2002}. He defined a weaker type of threshold (he calls {\em transition}), and provided a necessary and sufficient condition
for the existence of a transition, called {\em very
well-behavedness of the constraint set}.

\begin{definition}\label{csp-bad}
A value $\delta$ is {\em 0-bad} if there exists some canonical variable $x_{i}$ and constraint $C\in {\cal C}$ such that $C\models (x_{i}\neq \delta)$.
We say that $\delta$ is {\em $j$-bad} if there exists some constraint $C\in {\cal C}$ and variable $x_{i}$ such that $C\AND (x_{i}= \delta)$ implies that
some other variable $x_{k}$ of $C$ must be assigned a $j^{\prime}$-bad value, for some $j^{\prime}<j$. $\delta$ is {\em bad} if it is $j$-bad for some $j$
and {\em good} otherwise.
\end{definition}

\begin{definition}\label{csp-vwbehaved}
A set of constraints ${\cal C}$ is {\em well-behaved} iff:
\begin{enumerate}
\item there exists at least one good value in ${\cal D}$.
\item for every $\delta \in {\cal D}$ there exists at least one constraint $C\in {\cal C}$ not satisfied by the
assignment $(\delta, \delta, \ldots, \delta)$.
\end{enumerate}

${\cal C}$ is {\em very well-behaved} if, in addition to the
previous two properties, satisfies the following property: any
constraint formula from $CSP({\cal C})$ whose constraint
hypergraph is a cycle has a satisfying assignment where no
variable is assigned a bad value.
\end{definition}

Molloy has proved that the condition that ${\cal C}$ is very well behaved
is {\em necessary} for $CSP({\cal C})$ to have a sharp threshold.
Unfortunately this condition is not also sufficient for the existence of a
{\bf sharp threshold}: there exist \cite{molloy-stoc2002} pathological
examples of very-well behaved binary constraint satisfaction problems with a
4-ary domain that have a transition but do not have a sharp threshold:

\begin{example}\label{example-molloy}
Let ${\cal C}$ consist of two binary constraints over domain $\{0,1,2,3\}$. The first
constraint $C_{1}(x,y)$, forbids the pair $(x,y)$ from taking values from
the set $(0,0),(1,1), (2,2),(3,3)$. The second one, $C_{2}(x,y)$ forbids the situation when one of the variables takes a value from the set $\{0,1\}$ while the other constraint takes a value from the set $\{2,3\}$. Molloy's model allows for constraint templates to be applied with nonuniform probabilities, in this case $P(C_{1})=1/3$. $P(C_{2})=2/3$.
\end{example}

Molloy's counterexample involves non-uniform probabilities, so it would be
tempting to conjecture\footnote{we had an incorrect proof of this claim in an earlier version of the paper} that at least in the case of uniform probabilities very well-behavedness is necessary and sufficient for the existence of a sharp threshold of $CSP({\cal C})$. Unfortunately not even this is true\footnote{ we thank an anonymous referee for this counterexample.}:

\begin{example}\label{example-uniform}
Let ${\cal K}$ consist of three constraints of arity 4 over domain $\{0,1,2,3\}$ defined as follows (with respect to the constraints in Example~\ref{example-molloy}):
\[
K_{1}(x,y,z)= C_{1}(x,y),
\]
\[
K_{2}(x,y,z)=C_{2}(x,y),
\]
and
\[
K_{3}(x,y,z)=C_{2}(z,y),
\]
The constraints are applied with uniform probability. It is easy to see that
$CSP(K)$ is equivalent to the constraint satisfaction problem from
Molloy's counterexample~\ref{example-molloy}.
\end{example}

Nevertheless, in what follows we will obtain a necessary condition
for the existence of a sharp threshold that completely solves the
problem in the case of binary constraints applied with uniform
probability.

\begin{definition}\label{csp-ewbehaved}
A set of constraints ${\cal C}$ is {\em extremely well-behaved} iff:
\begin{enumerate}
\item ${\cal C}$ is very-well behaved.

\item There exists a  mapping $\Gamma$ from good values to
constraints in ${\cal C}$ such that, for every good value
$\delta$, constraint $\Gamma_{\delta} := \Gamma(\delta)\in {\cal
C}$ satisfies
\begin{equation}\label{extreme}
\Gamma_{\delta}(x_{1}, \ldots, x_{k})\vDash (x_{1}=\delta)\OR
\ldots \OR (x_{k}=\delta).
\end{equation}
\end{enumerate}
\end{definition}

\begin{theorem}\label{csp-dichotomy-threshold}
Let ${\cal C}$ be a set of extremely well-behaved constraints. Then $CSP({\cal C})$ has a sharp threshold.
\end{theorem}

The result of the theorem is incomparable with Molloy's sufficient condition for the existence of a sharp threshold: our result
does not imply the fact that 3-coloring has a sharp threshold \cite{achlioptas:friedgut:kcol}, while his does. On the other hand his result is not strong enough to yield the Corollary below.

We now apply the previous result to the case of boolean constraint
satisfaction (satisfiability) problems:

\begin{definition}
Constraint $C_{2}$ is {\em an implicate of $C_{1}$}
 iff every satisfying assignment for $C_{1}$ satisfies $C_{2}$.
\end{definition}

\begin{definition}
A boolean constraint $C$ {\em strongly depends on a literal} if it
has an unit clause as an implicate.
\end{definition}

\begin{definition}
A boolean constraint $C$ {\em strongly depends on a 2-XOR
relation} if $\exists i,j\in \{1,\ldots,k\}$ such that constraint
``$x_{i}\neq x_{j}$'' is an implicate of $C$.
\end{definition}

In this case we obtain the following explicit result, that totally
settles the case of random satisfiability problems, thus solving the open problem from \cite{creignou:daude:sat2002} and extending the results
from \cite{creignou-daude-thresholds}, where the result was shown
to hold under the additional restriction that all constraint
templates in set ${\cal P}$ are {\em symmetric}.

\begin{corollary}\label{dichotomy-threshold}
Consider a generalized satisfiability problem $SAT({\cal C})$
(that is not trivially satisfiable by the ``all zeros'' or ``all
ones'' assignment).
\begin{enumerate}
\item if some constraint in ${\cal C}$ strongly depends on one
component then $SAT({\cal C})$ has a coarse threshold. \item if
some constraint in ${\cal C}$ strongly depends on a 2XOR-relation
then $SAT({\cal C})$ has a coarse threshold. \item in all other
cases $SAT({\cal C})$ has a sharp threshold.
\end{enumerate}
\end{corollary}

\begin{proof}
The first two cases were proved by Creignou and Daud\'{e} in \cite{creignou:daude:sat2002}. In the
third case first, it is easy to see that both values 0 and 1 are
good, since there is no 0-bad value (hence no bad value, otherwise
${\cal C}$ would strongly depend on one variable).

Since constraints in ${\cal C}$ are not 0/1-valid and their domain is boolean
it
follows that condition (\ref{extreme}) in the definition of extreme
well-behavedness is satisfied. So all it is left to show is that
${\cal C}$ is very well behaved.

Indeed, consider a formula $\Phi$ whose associated hypergraph is a
cycle. If $\Phi$ were unsatisfiable then it would be minimally
unsatisfiable (since all acyclic formulas are satisfiable). But
Theorem 4.3 in \cite{creignou:daude:sat2002} prevents that from
happening, since for all minimally unsatisfiable formulas
$|Var(S)|\leq (k-1)|S|-1$, whereas for a cycle $|Var(S)(k-1)|S|$.

\end{proof}
\qedbox

\section{Proof of Theorem~\ref{csp-dichotomy-threshold}}

Our proof relies on the Friedgut-Bourgain criterion for the existence of a sharp threshold in {\em any} monotonic graph (or hypergraph) property $A$.

It is a well-known (and easy to see) fact that if property $A$ has a coarse threshold then there exists $0< \epsilon < 1/2$, $p^{*}=p^{*}(n)\in [p_{1-\epsilon}, p_{\epsilon}]$ and $C>0$ such that

\begin{equation}\label{condition-sharp}
p \cdot \frac{d\mu_{p}(A)}{dp}|_{p = p^{*}(n)} < C.
\end{equation}

Bourgain and Friedgut show\footnote{in \cite{friedgut:k:sat} the proposition is stated assuming for convenience that $p=p_{1/2}$, but this is not needed. We give here the general statement.}
 that

\begin{proposition} \label{sufficient-sharp}
Suppose $p= o(1)$ is such that condition~\ref{condition-sharp} holds
Then there is $\delta = \delta(C)>0$ such that either
\begin{equation}\label{first-condition-sharp}
\mu_{p}(x\in \{0,1\}^{n} | \mbox{ }x\mbox{ contains }x^{\prime}\in A\mbox{ of size }|x^{\prime}|\leq 10C\} > \delta
\end{equation}

or there exists $x^{\prime}\not \in A$ of size $|x^{\prime}| \leq 10C$ such that the conditional probability

\begin{equation}\label{second-condition-sharp}
 \mu_{p}(x\in A| x \supset x^{\prime})>\mu_{p}(A)+ \delta.
\end{equation}

\end{proposition}

\begin{observation}\label{obs-friedgut-survey}
We will, in fact, need one property that is {\em not} directly guaranteed by the Bourgain-Friedgut result as stated in \cite{friedgut:k:sat}, but follows from
an observation made in \cite{friedgut-survey-sharp}. For a finite set of words
$W$ define {\em the filter generated by $W$}, $F(W)$ as
\[
F(W)=\{x\mbox{ }|\mbox{ }(\exists y \in W)\mbox{ with }x\supseteq y\}.
\]

Friedgut observed (\cite{friedgut-survey-sharp}, remarks on pages
5-6 of that paper) that the statement of
condition~(\ref{second-condition-sharp}) can be strengthened in
the sense that the set $W$ of ``booster'' sets $x^{\prime}$
satisfies $\mu_{p}(F(W))=\Omega(1)$. Returning now to the case of
random constraint satisfaction problems, since the number of
isomorphism types of formulas of bounded size is finite, there
exists $\delta
>0$ and a satisfiable booster formula $F$ such that
\begin{enumerate}
\item condition (\ref{second-condition-sharp}) holds with $x^{\prime}=F$.
\item Formula $F$ appears with probability $\Omega(1)$ as a subformula in a random formula in $CSP_{p}({\cal C})$.
\end{enumerate}
\end{observation}

A standard observation is that in the second condition of    
Proposition~\ref{sufficient-sharp}, instead on conditioning on the presence 
of $x^{\prime}$ as a subset of $x$ one can, instead, add it: 

\begin{proposition} \label{sufficient-sharp-2} 
Suppose $p= o(1)$ is such that condition~(\ref{condition-sharp}) holds. 
Then there is $\delta = \delta(C)>0$ such that either
\begin{equation}\label{first-condition-sharp-2}
\mu_{p}(x\in \{0,1\}^{n} | \mbox{ }x\mbox{ contains }x^{\prime}\in A\mbox{ of size }|x^{\prime}|\leq 10C\} > \delta
\end{equation}

or there exists $x^{\prime}\not \in A$ of size $|x^{\prime}| \leq 10C$ such that
\begin{equation}\label{second-condition-sharp-2}
 \mu_{p}(x\cup x^{\prime}\in A)>\mu_{p}(A)+ \delta.
\end{equation}

\end{proposition}

Finally, note that for random constraint satisfaction problems, because of 
the invariance of such problems under variable renaming, one only needs to 
add a random copy of $x^{\prime}$. That is, the following 
version of Proposition~\ref{sufficient-sharp-2} holds:

\begin{proposition} \label{sufficient-sharp-3} 
Suppose $A=\overline{CSP({\cal C})}$ and $p= o(1)$ is such that 
condition~(\ref{condition-sharp})holds. 
Then there is $\delta = \delta(C)>0$ such that either
\begin{equation}\label{first-condition-sharp-3}
\mu_{p}(x\in \{0,1\}^{n} | \mbox{ }x\mbox{ contains }x^{\prime}\in A\mbox{ of size }|x^{\prime}|\leq 10C\} > \delta
\end{equation}

or there exists $x^{\prime}\not \in A$ of size $|x^{\prime}| \leq 10C$ such that, if $\Xi$ denotes the formula obtained by creating a copy of $x^{\prime}$ on a 
random tuple of variables, then 

\begin{equation} 
\label{second-condition-sharp-3}
 \mu_{p}(x\cup \Xi \in A)>\mu_{p}(A)+ \delta.  
\end{equation}

\end{proposition}

To show that random $CSP({\cal C})$  has a sharp threshold, we will
reason by contradiction. Assuming this is not the case,
one needs to prove that the
two conditions in Proposition~\ref{sufficient-sharp-3} do not hold. 

Suppose, indeed, that the first condition was true: that is, with positive probability it is true that a random formula $\Phi\in CSP({\cal C})$ contains
some unsatisfiable subformula $\Phi^{\prime}$ of size at most $10C$.
One can, therefore, assume that with positive probability  $\Phi$ contains a {\em minimally unsatisfiable formula} $\Phi^{\prime}$ of size at most $10C$.

On the other hand, with high probability all subformulas of a random formula
$\Phi$ of size at most $10C$ are either tree-like or unicyclic.
But because of well-behavedness, all formulas in $CSP({\cal C})$ that
are tree-like or unicyclic are satisfiable. Therefore the first condition
in Proposition~\ref{sufficient-sharp-3} cannot be true.

Assume, now, that the second condition is true: there exists a satisfiable
formula $F$ of size at most $10C$, such that conditioning on a random
formula in $CSP_{p}({\cal C})$ containing a copy of $F$ boosts the probability
of unsatisfiability by a value bounded away from zero. Because of the two conditions in Observation~\ref{obs-friedgut-survey} we infer that there exists a constant $\delta >0$ such that adding $F$ to a random
formula $\Phi\in CSP_{p}({\cal C})$ boosts the probability of
{\em un}satisfiability of the resulting formula by at least $\delta$.
As we discussed, we assume
that $F$ occurs with probability $\Omega(1)$ in a random formula in $CSP_{p}({\cal C})$. Therefore $F$ is tree-like or unicyclic (this argument was
also used in \cite{molloy-stoc2002}).

\begin{definition}
A {\em unit constraint} is a constraint (not necessarily part of the constraint set ${\cal C}$) specified by a condition $X=\delta$, with $X$ being a variable and
$\delta \in D$.
\end{definition}

\begin{lemma} \label{tree-sat}
Every tree-like or unicyclic formula
has a satisfying assignment $W$ consisting only of {\em good}
values.
\end{lemma}
\begin{proof}

This is easily proved by induction on the number of
clauses for tree-like formulas, even in a stronger form: if we set
one of the variables to an arbitrary good value, we can still set
the other variables to good values in such a way that we obtain a
satisfying assignment.

For a unicyclic formula we first set the variables appearing in
its unique cycle to good values so that all such
constraints are satisfied (this is possible since ${\cal C}$ is
very well-behaved). We are now left with several {\em tendrils},
tree-like formulas on disjoint set of variables, the root of each
such formula (the node appearing in the cycle) being set to a
fixed good value, which we can satisfy as in the first case.
\end{proof}
\qed

\begin{claim}
If $\Xi$ satisfies condition ~(\ref{second-condition-sharp-3}) then there exists another formula $G$ 
that is specified by a finite conjunction of unit constraints
\[
G\equiv (X_{1}=\delta_{1})\AND \ldots \AND (X_{p}=\delta_{p}),
\]

with all the values $\delta_{1}, \ldots, \delta_{p}\in {\cal D}$ being {\em good} values, and that also satisfies condition~(\ref{second-condition-sharp-3}). 
\end{claim}

\begin{proof}
Formula $\Xi$ appears with constant probability in a random $CSP({\cal C})$
formula with probability $p$ and has constant size. Therefore $\Xi$ is either
tree-like or unicyclic. The result follows easily from Lemma~\ref{tree-sat}, by replacing $F$ with formula $G$ consisting of the conjunction of unit constraints corresponding to a
satisfying assignment of $\Xi$ with good values. Indeed, $G$ is tighter than 
$\Xi$, so adding a random copy of $G$ instead of a random 
copy of $\Xi$ can only increase the probability that the resulting formula is unsatisfiable. 
\end{proof}
\qed

The key to refuting condition~(\ref{second-condition-sharp-3}) is to
show that, if it did hold then, for every monotonically increasing
function $f(n)$ that tends to infinity, we can also increase the
probability of unsatisfiability by a positive constant if, instead
of conditioning on $x$ containing a copy of formula $F$  we add
$f(n)$ random constraints from constraint set ${\cal C}$. We first
prove:

\begin{claim}\label{connecting-sharp-2}

Let $0<\tau < 1$ be a constant and let $p$ be such that $\mu_{p}(CSP({\cal C}))\geq \tau$. Assume that $r\geq 1$ and that $g_{1}, g_{2}, \ldots g_{r}$ are good values such that, when $(X_{1}, X_{2}, \ldots, X_{r})$ is a random
$r$-tuple of different variables

\begin{equation}\label {hypothesis-connecting-2}
Pr(\Phi\mbox{ has a satisfying assignment with } X_{1}=g_{1}, \ldots, X_{r}=g_{r}) \leq \frac{\tau}{2} .
\end{equation}

Then there exists constant $m\geq 1$ (that only depends on $k,r,\tau$) such that, if $\eta$ denotes a formula from $CSP({\cal C})$  obtained by adding,
for each good value $x$,  $m\cdot r\cdot 2^{k^{r}}$ random copies of $\Gamma(x)$, then
\begin{equation} \label{third-condition-sharp-2}
Pr(\Phi \cup \eta \mbox{ is satisfiable}) \leq \frac{\tau}{2}
\end{equation}
\end{claim}

\begin{proof}
We will give a proof of Claim~\ref{connecting-sharp-2} that is very similar
to that of the corresponding proof in \cite{achlioptas:friedgut:kcol}, thus
obtaining the desired contradiction.

For $i\in \{1,\ldots, r\}$ define $A_{i}$ to be the event that
the formula $\Phi$ has {\em no} satisfying 
assignment with the first  $i$ constraints $X_{1}=g_{1}, \ldots, X_{i}=g_{i}$
holding. Also define $A_{0}$ to be 
the event that $\Phi$ is {\em not} satisfiable.

The hypothesis translates as the fact that both inequalities $Pr(\overline{A_{0}})\geq \tau$ and \\
$Pr(\overline{A_{r}}) \leq \frac{\tau}{2}$ are true. Therefore 
\[
Pr(\overline{A_{r}}|\overline{A_{0}})= \frac{Pr(\overline{A_{r}}\AND \overline{A_{0}})}{Pr(\overline{A_{0}})}\leq \frac{\tau/2}{\tau} =\frac{1}{2}.
\] 
 
Thus we have 
\begin{equation}\label{initial-ineq}
\alpha_{r} := Pr[A_{r}|\overline{A_{0}}]= Pr[A_{r-1}|\overline{A_{0}}]+Pr[A_{r}|\overline{A_{r-1}}\AND \overline{A_{0}}]\cdot Pr[\overline{A_{r-1}}|\overline{A_{0}}]\geq \frac{1}{2}
\end{equation}

$Pr[A_{r}|\overline{A_{r-1}}\AND \overline{A_{0}}]= Pr[A_{r}|\overline{A_{r-1}}]$ is the fraction of variables that have to receive values different from $g_{r}$ if constraints $X_{1}=g_{1}, \ldots, X_{r-1}=g_{r-1}$ are added to $\Phi$; let $C_{r}$ be the set of such variables.
If instead of the last constraint we add a random copy of the constraint $\Gamma(g_{r})$ we spoil satisfiability as well when all the variables appearing in the constraint are in the set $C_{r}$. Denoting $\lambda_{r}= (Pr[A_{r}|\overline{A_{r-1}}])^{k}$, this last event happens with probability
$\lambda_{r}/(1-o(1))$, so the probability that the resulting random formula is unsatisfiable is at least 
\begin{equation}
\beta_{r}:= Pr[A_{r-1}|\overline{A_{0}}]+\frac{\lambda_{r}}{1-o(1)}\cdot Pr[\overline{A_{r-1}}|\overline{A_{0}}].
\end{equation}

Because of the convexity of the function $f(x)=x^{k}$ and constraint~\ref{initial-ineq}, by applying Jensen's inequality it follows that
\begin{eqnarray*}
\frac{1}{2^k} & \leq & \alpha_{r}^{k}= (Pr[A_{r-1}|\overline{A_{0}}]\cdot 1+Pr[A_{r}|\overline{A_{r-1}}]\cdot Pr[\overline{A_{r-1}}|\overline{A_{0}}])^{k} \leq \\ & \leq & Pr[A_{r-1}|\overline{A_{0}}]\cdot 1^{k}+Pr[A_{r}|\overline{A_{r-1}}]^{k}\cdot Pr[\overline{A_{r-1}}|\overline{A_{0}}] = \\ & = & (Pr[A_{r-1}|\overline{A_{0}}] +\lambda_{r}\cdot Pr[\overline{A_{r-1}}|\overline{A_{0}}])=\beta_{r}\cdot (1+o(1)).
\end{eqnarray*}

Thus $\beta_{r}\geq \frac{1}{2^k}\cdot (1-o(1))$. The conclusion of this argument is that adding one random copy of $\Gamma(b_{r})$ instead of the $r$-th constraint lowers the probability of unsatisfiability to no less than  $\frac{1}{2^k}\cdot (1-o(1))$. Adding the copy of the constraint {\em before} the first $r-1$
constraints and repeating the argument recursively implies the fact that, if
instead of adding the $r$ constraints to $\Phi$ we add $r$ random copies
of $\Gamma(b_{1}), \ldots, \Gamma(b_{r})$ the probability of unsatisfiability
of the resulting formula, given that $\Phi$ was satisfiable, is at least $\gamma_{r} = \frac{1}{2^{k^{r}}(1-o(1))}$. Since the values $b_{1}, \ldots, b_{r}$ can repeat themselves, the same is true if we add $r$ random copies of $\Gamma(x)$ for {\em every} good value $x$.

Suppose now that we add $r\cdot m\cdot 2^{k^{r}}$ copies of each
$\Gamma(x)$ instead (that is, we repeat the random experiment
$m\cdot 2^{k^{r}}$ times, for some integer $m\geq 1$). By doing so
the probability that none of the experiments will make the resulting formula 
unsatisfiable is at most $(1-\gamma_{r})^{m\cdot 2^{k^{r}}}$. For
some constant $m$ this is going to be at most $1-\frac{\tau}{2}$.
This means that

\begin{equation}
Pr(\Phi \cup \eta \mbox{ is satisfiable}) \leq \frac{\tau}{2}
\end{equation}
\end{proof}
\qed

Condition~\ref{third-condition-sharp-2} can be refuted directly, leading to a contradiction. This is done e.g. by Lemma 3.1 in \cite{achlioptas:friedgut:kcol}. For convenience, we now restate this result:

\begin{lemma}\label{achlioptas:friedgut} For a monotone property
\footnote{Achlioptas and Friedgut assume $A$ to be a monotone {\em graph}
property, but this fact is not used anywhere in their proof.}  $A$ let
\[
\mu(p)= Pr[G\in \Gamma(n,p)\mbox{ has property }A],
\]
\[
\mu^{+}(p,M)=Pr[G_{1}\cup G_{2}\mbox{ }|\mbox{ }G_{1}\in \Gamma(n,p), G_{2}\in \Gamma(n,M)\mbox{ has property }A].
\]

Let $A=A(n)\subseteq \{0,1\}^{n}$ be a monotone property and $M=M(n)$ such that  $M= o(\sqrt{np})$. Then:
\[
|\mu(p)-\mu^{+}(p,M)|= o(1).
\]
\end{lemma}

Now it is easy to obtain a contradiction: consider a random formula $\eta$
with $f(n)$ clauses, for some $f(n)\goesto \infty$.
It is easy to show that the probability that $\eta$
contains, for some good value $x$, less than $M$ copies of constraint $\Gamma(x)$ is $o(1)$. So adding $\eta$ instead of the random formula in Claim~\ref{connecting-sharp-2} only increases the probability of the resulting formula being unsatisfiable. But
then the conclusion of Lemma~\ref{achlioptas:friedgut} directly contradicts
that of Claim~\ref{connecting-sharp-2}.



\section{Conclusions}

What made the proof work ? Its main steps are, of course, quite
similar to the proofs of similar results in \cite{friedgut:k:sat,achlioptas:friedgut:kcol,istrate:ccc00,molloy-stoc2002}. One element we want to highlight is a certain {\em monotonicity} property used in the proof of  Claim~\ref{connecting-sharp-2}, since it already proved useful in obtaining further insights on classifying threshold properties of random constraint satisfaction problems \cite{istrate:ccc00,istrate:descriptive}, and can reasonably be expected to help in obtaining a complete classification: In Claim~\ref{connecting-sharp-2} we have only used the fact that for every good value $\delta_{1}$ there exists a constraint $C_{\delta_{1}}(x_{1},\ldots, X_{k})$ in the image of map $\Gamma$ that implies the constraint $(X_{1}=\delta_{1})\OR \ldots \OR  (X_{k}=\delta_{1})$. This means that as long as the other steps  of the proof continued to work we could have proved the Claim if the constraint $C_{\delta_{1}}$ was {\em exactly} the constraint $(X_{1}=\delta_{1})\OR \ldots \OR  (X_{k}=\delta_{1})$. This provides a general strategy
for proving sharp threshold results:
\begin{enumerate}
\item identify a ``base case'' $B$ for which the analog of Claim~\ref{connecting-sharp-2} can be proved.
\item If we are given a set of constraints ${\cal C}$ that are tighter than the ones of the base case $B$, to prove that  $CSP({\cal C})$ has a sharp threshold it is enough to verify that all the other steps of the proof still hold. The
analog of Claim~\ref{connecting-sharp-2} will now follow from the corresponding claim for the base case $B$.
\end{enumerate}

To sum up, we have proved a sufficient condition for the existence of a sharp
threshold for random constraint satisfaction problems that
completely solves the boolean case. Example~\ref{example-uniform}
showed, however, that the results could not be extended to all
very-well behaved sets of constraints. Obtaining a necessary and
sufficient condition for the existence of a sharp threshold in
random constraint satisfaction problems is an interesting open
problem.

\bibliography{/home/gistrate/bib/bibtheory}

\begin{thebibliography}{10}

\bibitem{achlioptas:friedgut:kcol}
D.~Achlioptas and E.~Friedgut.
\newblock A sharp threshold for $k$-colorability.
\newblock {\em Random Structures and Algorithms}, 14(1):63--70, 1999.

\bibitem{bol:b:random-graphs}
B.~Bollob\'{a}s.
\newblock {\em Random Graphs}.
\newblock Academic Press, 1985.



\bibitem{creignou:daude:sat2002}
N.~Creignou and H.~Daud\'{e}.
\newblock Generalized satisfiability problems: minimal elements and phase
  transitions.
\newblock {\em Theoretical Computer Science}, 302(1--3):417--430, 2003.

\bibitem{creignou-daude-thresholds}
N.~Creignou and H.~Daud\'{e}.
\newblock Combinatorial sharpness criterion and phase transition classification
  for random {CSP}s.
\newblock {\em Information and Computation}, 190(2):220--238, 2004.


\bibitem{friedgut:k:sat}
E.~Friedgut.
\newblock Necessary and sufficient conditions for sharp thresholds of graph
  properties, and the k-{SAT} problem. with an appendix by {J}. {B}ourgain.
\newblock {\em Journal of the A.M.S.}, 12:1017--1054, 1999.

\bibitem{friedgut-survey-sharp}
E.~Friedgut.
\newblock Hunting for sharp thresholds.
\newblock {\em Random Structures and Algorithms (Early View article DOI
  10.1002/rsa.20042 )}, 2004.

\bibitem{istrate:ccc00}
G.~Istrate.
\newblock Computational complexity and phase transitions.
\newblock In {\em Proceedings of the 15th I.E.E.E. Annual Conference on
  Computational Complexity (CCC'00)}, 2000.

\bibitem{istrate:horn}
G.~Istrate.
\newblock The phase transition in random {H}orn satisfiability and its
  algorithmic implications.
\newblock {\em Random Structures and Algorithms}, 4:483--506, 2002.

\bibitem{istrate:descriptive}
G.~Istrate.
\newblock Descriptive complexity and first-order phase transitions.
\newblock (in progress).


\bibitem{martin:monasson:zecchina}
O.~Martin, R.~Monasson, and R.~Zecchina.
\newblock Statistical mechanics methods and phase transitions in combinatorial
  optimization problems.
\newblock {\em Theoretical Computer Science}, 265(1-2):3--67, 2001.

\bibitem{molloy-stoc2002}
M.~Molloy.
\newblock Models for random constraint satisfaction problems.
\newblock In {\em Proceedings of the 32nd ACM Symposium on Theory of
  Computing}, 2002.




\end{thebibliography}

\end{document}